# Infrared vertically-illuminated photodiode for chip alignment feedback

L. Alloatti[1,a)], R. J. Ram[1]

[1]Massachusetts Institute of Technology, Cambridge, MA 02139, USA

[a)]Current address: Institute of Electromagnetic Fields (IEF), ETH Zurich, Zurich, Switzerland

[a)]luca.alloatti@ief.ee.ethz.ch

We report on vertically-illuminated photodiodes fabricated in the GlobalFoundries 45nm 12SOI node and on a packaging concept for optically-interconnected chips. The photodiodes are responsive at 1180 nm −a wavelength currently used in chip-to-chip communications. They have further a wide field-of-view which enables chip-to-board positional feedback in chip-board assemblies. Monolithic integration enables on-chip processing of the positional data.

State-of-the-art electrical chips, such as the Titan GK110 graphic processing unit (GPU) of NVIDIA, require more than 40 Tbit/s I/O bandwidth based on the 1 byte per FLOP rule-of-thumb.[1,2] However, the bottleneck caused by electrical communications limits the available bandwidth of such chips to less than a tenth of their needs. For overcoming such limitations, future microprocessors and memories will likely communicate through high-bandwidth and energy-efficient optical links.[3,4,5,6]

The complex network topology of such systems is ideally defined by an optical substrate, or optical circuit board (OCB), containing arrays of single-mode waveguides, waveguide crossings, waveguide splitters and couplers. Such components have been demonstrated in a variety of material systems, including silicon nitride and polymers.[7,8,9,10,11] In these concepts the light can be coupled in and out of the chip through pairs of grating couplers, Fig. 1. However, the alignment tolerances between chip and OCB are dictated by the bandwidth of the grating couplers and are generally in the sub-$\mu$m range.[12] While commercial chip bonders with the required precision are readily available,[13] they generally require a positional feedback based on optical imaging, therefore limiting this technology to chips and/or substrates which are transparent in the visible or infrared range. Furthermore, no solution has been proposed so far for enabling end-users to plug components on the board as in current electrical systems.

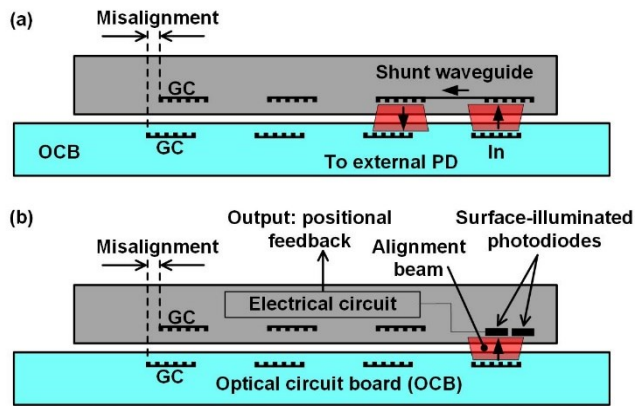

Fig. 1. A photonic chip is placed on an optical circuit board (OCB). Optical coupling is provided by pairs of grating couplers (GC) in the chip and OCB. (a) In the loop-back method, the misalignment between the grating couplers is estimated by measuring the insertion loss trough two pairs of grating couplers and a shunt waveguide in the chip. (b) The current proposal consists of vertically-illuminated photodiodes which collect the light emitted from a coupler in the OCB. The photocurrent is processed by on-chip electrical circuits which deliver positional feedback. The differential read-out of the photodiode enables sub-$\mu$m accuracies.

To overcome these limitations, previous work has relied on a pair of grating couplers directly connected together by a shunt waveguide on the chip side, Fig. 1(a), and to align the components by minimizing the total insertion loss (loop-back configuration).[14,15] This approach however does not instantaneously provide the direction in which the components need to be moved for achieving optimal coupling, requires two sacrificial external waveguides (in the OBC or in the fiber bundle) for each shunt waveguide, and differential read-out is not possible. In this scheme, moreover, the required initial alignment tolerances are dictated by the field-of-view of the grating couplers, and external power monitors need to be aligned prior to the start of the positioning process.

In this work, we demonstrate vertically-illuminated photodiodes suitable for chip-to-OCB positional feedback and propose a packaging concept for optically-interconnected chips. In this scheme, one or more grating couplers on the OCB emit a single-mode beam towards photodiode pairs located on the chip, Fig. 1(b). The differential signal on the photodiodes is read out by on-chip electrical circuits which process the positional information. The wavelength emitted from the gratings used for alignment should be close to the wavelength used for the optical links so that all waveguides in the OCB can have the same cross section (to ease fabrication) and remain single mode.

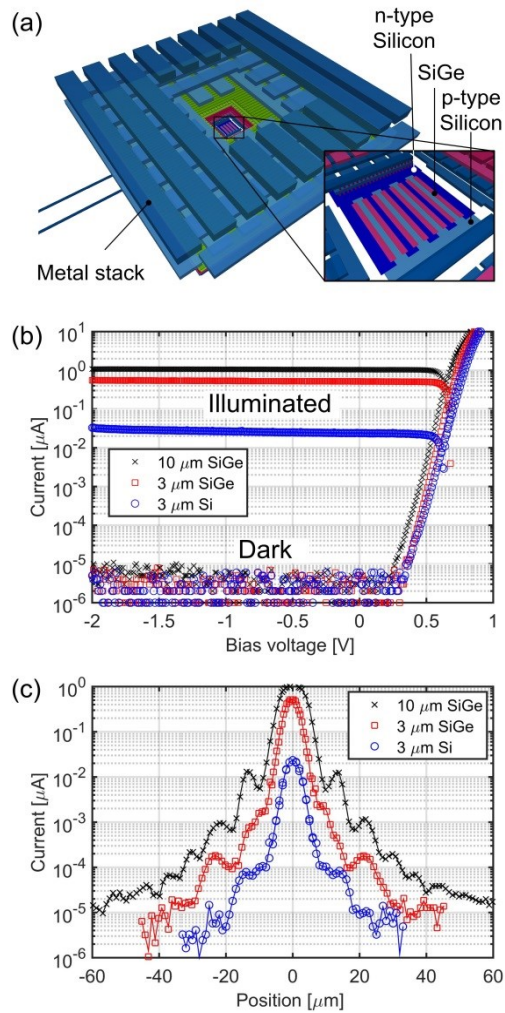

Fig. 2 Characteristics of the fabricated photodiodes. (a) Three-dimensional representation of a fabricated photodiode in the 45 nm 12SOI process. The light is coupled through metal apertures in the metal stack and the photodiode consists of silicon-germanium stripes contacted by interdigitated n-type and p-type silicon. (b) Dark and illuminated current-voltage characteristics of 10 $\mu$m × 10 $\mu$m SiGe, 3 $\mu$m × 3 $\mu$m SiGe or 3 $\mu$m × 3 $\mu$m silicon-only square photodiodes (legend). For the 10 $\mu$m × 10 $\mu$m photodiode a current of 1 $\mu$A is obtained, corresponding to a responsivity of $3.2 \times 10^{-5}$ A/W. (c) Current vs. position characteristics for the different photodiodes (legend) at 0 V bias. For all the data, the optical power is of 15 dBm at a wavelength of 1180 nm.

Test photodiodes have been fabricated in the GlobalFoundries (former IBM) 45nm 12SOI process, without requiring the modification of the fabrication flow and without violating the original design rules,[16] a concept that we named "zero-change CMOS".[17] In this node, we have recently demonstrated a complete toolbox of photonic components, including high-speed and high-responsivity photodiodes,[18,19] high-speed modulators[20] and low-loss grating couplers[21] monolithically integrated next to million-transistors circuits. Earlier waveguide versions of such devices enabled the first single-chip microprocessor communicating directly using light.[6] The optical components operate at a wavelength of 1180 nm for optimal responsivity of the photodiodes and for achieving a better confinement in the sub-100 nm thick crystalline silicon layer.

For increasing the responsivity of the surface-illuminated test photodiodes, we have exploited silicon-germanium (SiGe) stripes as active regions, inset in Fig. 2(a). The SiGe is normally utilized in the 45nm 12SOI process for increasing the carrier mobility in pFETs through compressive strain, is heteroepitaxially grown in silicon pockets and has an estimated germanium atomic content of 25% - 35%. Interdigitated n-well and p-well implants contact the SiGe stripes and act as charge collectors, Fig. 2(a). All photodiodes have a square cross-section and have size 10 $\mu$m × 10 $\mu$m or 3 $\mu$m × 3 $\mu$m. An additional 3 $\mu$m × 3 $\mu$m photodiode without SiGe has been included as reference. In our experiments, the photodiodes were illuminated with a cleaved SMF28 fiber located 5 $\mu$m to 10 $\mu$m above the chip surface and forming an angle of 8° with the normal. The photodiodes are accessed through the back-end-of-line (BEOL) dielectric stack through apertures in the metal layers Fig. 2(a).

The current-voltage characteristics of the three devices are shown in Fig. 2(b) under 15 dBm illumination at a wavelength of 1180 nm. The presence of SiGe increases the responsivity by a factor of 22 as compared to only-silicon detectors at 0 V bias. The dark current is smaller than 10 pA for all devices in the -2 V to 0 V range. The dark current is significantly lower than those of germanium photodiodes demonstrated in modified processes, which is typically four orders of magnitude larger for similar device dimensions at a bias voltage of -1 V.[22] The small dark current is attributed to a low density of defects and dislocations.[19]

The photocurrent was then recorded as the fiber was moved along the direction parallel to the SiGe stripes, Fig. 2(b). For an optical power of 15 dBm and 0 V bias, the widest photodiode has a field-of-view of 40 $\mu$m at a 1 nA current threshold and 120 $\mu$m at a 10 pA threshold.

In the remaining of the paper we describe a packaging concept which takes advantage of such photodiodes and we will assume that the overall mechanical precision of the packaging assembly (which defines the initial alignment tolerances) is better than 100 $\mu$m. A possible configuration of the alignment-photodiodes is shown in Fig. 3(a). Here, four 10 $\mu$m × 10 $\mu$m SiGe photodiodes are placed on the vertices of a square with 40 $\mu$m edges so that the system would generate a current larger than 1 nA for alignment inaccuracies of up to ±40 $\mu$m (or 10 pA for ±100 $\mu$m). By placing 3 $\mu$m × 3 $\mu$m SiGe photodiodes at the target beam position as shown in Fig. 3(a) and using the data in Fig. 2(c) the differential photocurrent would be of about 30 nA per 100 nm of displacement. Such a set of photodiodes would therefore provide a field-of-view larger than the mechanical precision of the package and a sufficient sensitivity for obtaining the required sub-$\mu$m tolerances. Three such sets of photodiodes are placed around the area occupied by the grating couplers, Fig. 3(b), such that the translation and rotation required for achieving optimal chip-to-OCB coupling can be determined. The current generated by the 32 photodiodes is processed by monolithically integrated circuits and the resulting signal is used either for feedback to chip-bonders for a one-time alignment, or to control mechanical actuators of an optical socket as shown in Fig. 4.

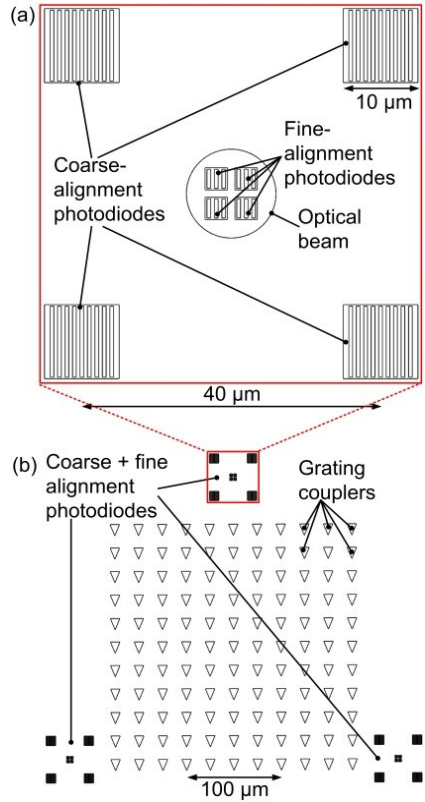

Fig. 3. Representation of the grating couplers and of the alignment photodiodes. (a) Representation of alignment structures consisting of four large-area planar silicon-germanium photodiodes for coarse-alignment and four small-area photodiodes for fine alignment. The spot size of the alignment beam is represented for reference at the position of ideal alignment. (b) An array of 121 on-chip grating couplers (waveguides not shown) is used as optical interface to the OCB. At the periphery of the grating array three alignment structures are located.

In a possible embodiment of an optical socket, optical coupling between the chip and the OCB occurs near one of the vertices of the chip, Fig. 4. The chip is mounted with the grating couplers facing downwards and the metal stack on the opposite side. The chip is partially encapsulated (metal side covered, handle wafer side free as in standard CPUs) in a movable mount, made for example of ceramics, which is connected through flexible wires to a frame (outer ceramics) fixed to the board. Since future optically interconnected chips will likely transfer the majority or all the data through optical links, only a small number of wires will be required (for DC power, clock signals and minimal amount of data such as the positional information from the alignment photodiodes). On the top metal layer of the chip, the high-density electrical pads used for I/O in current microprocessors is replaced with a dense metal grid for distributing the ~100 A needed for power supply. An electrical printed circuit board (PCB) is located under the OCB. Both the chip and the OCB are realized with optical lithography which well guarantees sub-$\mu$m accuracy –this is a necessary condition for being able to align all the couplers at once. In the example shown in Fig. 4, the chip uses a silicon-on-insulator (SOI) platform where the silicon substrate (handle wafer) has been removed at the location of the gratings (partial substrate release).[17] In all other locations (where the electronics is present) the substrate is not removed and is in contact with a heatsink which passes through both the PCB and the OCB. The position of the inner mount is adjusted by micrometer positioners (such as piezoelectric actuators) which enable in-plane translation and rotation so that optimal alignment between the on-chip couplers and the OCB couplers is achieved. Such positioners are connected to a driver on the PCB which is directly controlled by the signal generated on the chip.

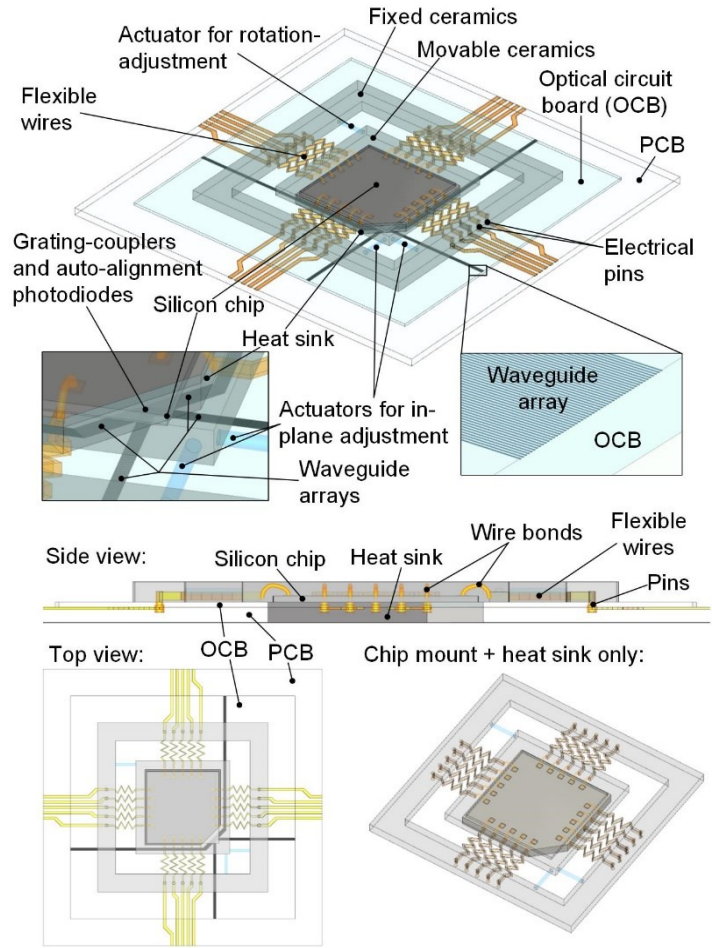

Fig. 4. Representation of an optical socket. The silicon chip is fixed in a movable ceramics with the electrical contacts on the top side, and a grating coupler array on the bottom side. This ceramics is suspended through flexible wires transporting electrical power, clock signal and minimal data. These wires are connected to an outer ceramics fixed on the board. Two linear actuators are in charge of moving the inner ceramics in the plane. A third actuator enables rotation adjustments. Piezoelectric actuators having a 10 mm in length, 100 V operation voltage and 10 $\mu$m travel are readily available. A screw (not shown) inserted in the external ceramics in series with each piezoelectric actuator enables a one-time rough alignment feed backed by the photodiodes for up to few hundreds of micrometers in travel. The silicon chip may consist of a silicon-on-insulator (SOI) die with released silicon substrate at the location of the grating coupler array.[6] The silicon body is in contact to the heat sink which passes through both the printed circuit board (PCB) and the optical circuit board (OCB).

In conclusion, we have presented vertically-illuminated photodiodes in a 45nm CMOS process for facilitating the packaging of optically-interconnected chips. The concept exploits the large field-of-view and the small dark-currents of the photodiodes for providing positional feedback to a mechanical alignment system. The positional feedback can either be used for a one-time alignment, or for creating optical sockets in modular systems. The approach exploits monolithic integration of electrical and optical components for handling the complexity of wiring a large number of photodiodes and for processing the positional information.

We acknowledge support by DARPA POEM under award HR0011-11-C-0100 and contract HR0011-11-9-0009. The views expressed are those of the authors and do not reflect the official policy or position of the DoD or the U.S. Government.